\documentclass[aps,twocolumn,nofootinbib]{revtex4}

\usepackage{times}
\usepackage{amsmath}
\usepackage{amsfonts}
\usepackage{amssymb}
\usepackage{graphicx}

\begin{document}

\title{Connections among three roads to cosmic acceleration: \\ decaying vacuum, bulk viscosity, and nonlinear fluids}
\author{Sandro Silva e Costa}
\email{sandro.costa@ufabc.edu.br}

\affiliation{Centro de Ci\^encias
Naturais e Humanas -- Universidade Federal do ABC\\
Rua Santa Ad\'elia, 166 -- 09210-170 \\ Santo Andr\'e -- SP -- Brazil\\
and\\
Departamento de F\'\i sica -- UFMT \\
Av.  Fernando  Correa da Costa, s/n$^o$ -- 78060-900 \\ Cuiab\'a
 --  MT -- Brazil}
\author{Mart\'in Makler}
\email{martin@cbpf.br} \affiliation{
Centro Brasileiro de Pesquisas F\'\i sicas 
\\
R. Dr. Xavier Sigaud, 170 -- 22290-180 \\ Rio de Janeiro -- RJ --
Brazil}

\date{\today}

\begin{abstract}
We discuss the connection among three distinct classes of models
often used to explain the late cosmic acceleration: decaying
cosmological term, bulk viscous pressure, and nonlinear fluids. We
focus on models that are equivalent at zeroth order, in the sense
they lead to the same solutions for the evolution of the scale
factor. More specifically, we show explicit examples where this
equivalence is manifest, which include some well know models
belonging to each class, such as a power law $\Lambda$-term, a model
with constant viscosity, and the Modified Chaplygin Gas. We also
obtain new analytic solutions for some of these models, including a
new \emph{Ansatz} for the cosmic term.

\end{abstract}

\maketitle

\section{Introduction}

The combination of a large set of astrophysical observations shows
that the Universe is currently undergoing a phase of accelerated
expansion. A natural explanation for this phenomenon, in the
framework of Einstein's theory of gravitation, would be the presence
of a cosmological constant. However, if this constant is interpreted
as the vacuum energy, a mismatch of about 50 to 100 orders of
magnitude occurs between the different contributions to the vacuum
energy and the observed value of $\Lambda$ (this is known as the
\emph{cosmological constant problem}). Besides, if the cosmological
term is constant, it implies that we live in a very special period
of cosmic history where the contribution of $\Lambda$ to the total
energy density is of the same order of magnitude of that of the
matter fields (this is the so-called \emph{coincidence problem}).

An interesting alternative, put forward by M. P. Bronstein already
in 1933 \cite{Bronstein}, is the possibility of a \emph{decaying
cosmological term}. Since then, several phenomenological models were
proposed for this component. Some authors argue that a dynamical
$\Lambda$ is even a requirement of Quantum Field Theory and provide
first principles estimations of its dependence on cosmic expansion
(see, e.g., refs. \cite{Adler82,ShapiroetalPLB03}). A suitable
$\Lambda$ decaying model could potentially connect the primordial
inflation with the present acceleration and also alleviate the
coincidence problem.

Another possibility as a driving force for the accelerated expansion
is the presence of an isotropic viscous pressure (see, e.g., refs.
\cite{Zimdahl2001,Balakin2003,BulkJulio}), which can arise from
dissipation processes 
or as a consequence of particle creation.

A third road to cosmic acceleration is provided by so-called
\emph{exotic fluids}, which have negative relativistic pressure,
such as the the widely used linear equation of state (EOS)
$p=w\rho$, with $w<-1/3$. However, a fluid with nonlinear EOS may
have more suitable properties, such as being stable with respect to
pressure perturbations. An example is given by the Chaplygin gas
\cite{Chaplygin,Kamen01}. A particularly interesting property of
this type of fluid is the possibility of unifying \emph{dark energy}
(DE) and \emph{dark matter} (DM) (for a review of this scenario,
see, e.g., ref. \cite{RevQuart}).

Many other frameworks which lead to an acceleration of the cosmic
expansion have been proposed, most notably those based on scalar
fields --- generically denoted as quintessence. For recent reviews
on DE, with special emphasis on quintessence, see, e.g., refs.
\cite{PeeblesRatra03,PadmanbhamReviewDE03,DEreviews}. It is worth
noting that all these approaches assume that the constant
contribution of vacuum energy plus cosmological constant are
identically zero due to some compensation mechanism and therefore do
not address the cosmological constant problem.

In this work we discuss the connections among the first three
possible explanations for the accelerated expansion mentioned above,
namely, decaying $\Lambda$, bulk viscous pressure, and exotic fluids
with nonlinear EOS. We show that these three classes of models
modify in the same way the evolution of the expansion rate.
Therefore for a given model in one class, it is possible to find
equivalent models belonging to the other classes that produce the
same expansion history of the Universe.

We discuss some explicit examples of this connection, displaying the
equivalent models for two popular sets of $\Lambda$ decaying
cosmologies: one where the cosmic term has a power law dependence on
the scale factor and another where it has a quadratic form in terms
of the expansion rate.

In the process of investigating the equivalent models, we derive
analytical solutions for the scale factor in some particular cases.
We also introduce and find analytical solutions for a new model of
decaying $\Lambda$ term. These solutions are useful to test
numerical codes, they provide some insight for more generic
numerical solutions, and are handy for pedagogical purposes.

A corollary of the discussion presented in this work is that one
cannot observationally distinguish these three possible explanations
for the accelerated expansion using only background results (i.e.,
the zeroth order homogeneous evolution of the Universe), such as the
supernovae distance-redshift relation.

The paper is organized as follows. In section \ref{3types} we
present a brief review of the three frameworks to explain the
accelerated expansion on which we focus in this work, making
apparent the connections among them. Some specific examples are
discussed in section \ref{results}, where equivalent models are
explicitly shown and analytic solutions are found. We summarize our
results in section \ref{summary}. Finally, we present our concluding
remarks, and discuss avenues of future research in section
\ref{discussion}.

Throughout the text we use natural units, where $c=G=\hbar=k_B=1$.

\section{Three possibilities for cosmic acceleration\label{3types}}

In this section we briefly review the three roads to cosmic
acceleration and present some models that will be discussed
throughout this work. We introduce the basic equations that govern
the evolution of the scale factor in these frameworks, exhibiting
the formal analogy between them.

\subsection{Decaying cosmological term \label{Deacaying}}

A dynamic cosmological term (i.e., a component with energy-momentum
tensor $T_{\mu\nu}=\Lambda\left(  x\right)  g_{\mu\nu}$) emerges
naturally in the framework of quantum field theory (see, e.g., ref.
\cite{Adler82} and refs. therein). Cosmological models with
time-varying $\Lambda$ were introduced in the 1980's (see, e.g.,
refs.
\cite{Bertolami86,OzerTaha,Fresse87,Wetterich,Gasperini,Peebles88})
either as alternatives to the inflationary paradigm and to a
primordial singularity, or to reconcile inflation with observational
data. Such $\Lambda$ decaying cosmologies attracted more interest in
the 1990's, also in connection with the age problem, and more models
where proposed and tested against the available data (see, e.g.,
refs.
\cite{ChenWu90,Pavon91,WagaPower,CarLimWa1992,Waga93,SalimWaga93,LimaMaia}).
Finally, this type of model became popular in the turn of the
century, impelled by supernovae and cosmic microwave background
(CMB) evidence for cosmic acceleration and a smoothly distributed
DE.

In most $\Lambda$ decaying models, the density associated to the cosmological
term is either determined by the dynamics of a scalar field or is given as an
explicit expression in terms of the scale factor $a$ and/or the expansion
factor $H$. Here we shall consider only the later case. For example, \"{O}zer
and Taha \cite{OzerTaha} introduced a model in which $\Lambda\propto a^{-2}$.
This same dependence was proposed by Chen and Wu \cite{ChenWu90}, based on
dimensional arguments in the context of quantum cosmology (see also
\cite{Pavon91,LopezNanopoulos96}). Gasperini, using the thermal interpretation
of $\Lambda$ \cite{GibbonsHawking}, introduced the more general power law form
\cite{Gasperini}%
\begin{equation}
\Lambda\propto a^{-m}. \label{Lpower}%
\end{equation}
Several observational consequences of this model were investigated
by Waga and collaborators \cite{WagaPower} and it was found to be
consistent with the existing data for $m\gtrsim1.6$. More recently,
models in which the energy density associated to the cosmic term has
the form
\[
\rho_{\Lambda}=\rho_{\Lambda0}+\frac{\varepsilon\rho_{m0}}{3-\varepsilon}a^{-3+\varepsilon}
\]
where also proposed \cite{ShapiroetalPLB03,WangMeng} and compared
with observational data \cite{AlcanizLimaPRD72,OpherPelinsonPRD04}.
In particular, a comparison of Large Scale Structure (LSS) data with
CMB observations leads to the constraint $\varepsilon\lesssim
0.7\times10^{-3}$ \cite{OpherPelinsonPRD04}.

Models in which $\Lambda$ has a scaling in terms of the expansion
factor were also proposed. For example, investigations using the
renormalization group approach lead to a running of the cosmic term
of the form\footnote{A quadratic scaling was also introduced
previously, motivated by dimensional arguments
\cite{CarLimWa1992}.} \cite{ShapiroetalPLB03,BonetetalJCAP04,ShaSoSteJCAP05}%
\begin{equation}
\Lambda=\Lambda_{0}+\nu H^{2} \, . \label{H2}%
\end{equation}
This model is consistent with a number of observational data, but
again a comparison with LSS and CMB data implies
$\nu\lesssim2.3\times10^{-3}$ \cite{OpherPelinsonPRD04} (see also
\cite{AlcanizLimaPRD72}). On the other hand, an estimate of the
cosmological term from the trace anomaly of quantum
chromodynamics (QCD) yields \cite{Schutzhold02}%
\begin{equation}
\Lambda=\sigma H \, , \label{Hlin}%
\end{equation}
where the proportionality constant would be related to the QCD cutoff energy
scale ($\sigma=\Lambda_{QCD}^{3}$).

Several other models, with different phenomenological scalings in
$H$, $a$, or a combination of the two, have also been introduced
(see, for instance, refs.%
\footnote{See also ref. \cite{OverduinCooperstock98}, for a
collection of several phenomenological $\Lambda$-decay laws and some
historical remarks.}
\cite{CarLimWa1992,Waga93,SalimWaga93,LimaMaia,Bauer05}). However,
in what refers to specific models, in the remaining of the paper we
shall restrict to the simple scalings given by eqs. (\ref{Lpower}),
(\ref{H2}), and (\ref{Hlin}).

Now, let us state the basic equations that govern the dynamics of
models with a cosmic fluid and a cosmological term (in a
Friedmann--Lema\^itre--Robertson--Walker Universe). This will allow
to make the equivalence with the other two frameworks explicit.
Since a component with $p_{\Lambda}=-\rho_{\Lambda}$ can only be
dynamic if it interacts with another matter-energy component, we
shall impose only that the total energy-momentum is conserved,
leading to the energy conservation equation%
\begin{equation}
\dot{\rho}+3H\left(  p+\rho\right)  =-\dot{\rho}_{\Lambda} \,, \label{econsL}%
\end{equation}
where $H=\dot{a}/a$ and the dot denotes derivative with respect to
the cosmic time $t$. Differentiating the Friedmann equation

\begin{equation}
H^{2}=\frac{8\pi}{3}\left(  \rho+\rho_{\Lambda}\right)  -\frac{k}{a^{2}}
\label{Fried}%
\end{equation}
and using equation (\ref{econsL}) leads to%
\begin{equation}
\dot{H}-\frac{k}{a^{2}}=-4\pi\left(  p+\rho\right) \, . \label{FriedDot}%
\end{equation}
From now on, we shall consider only the case in which the cosmic
fluid has a linear
barotropic equation of state:%
\begin{equation}
p=\left(  \gamma-1\right)  \rho\,. \label{plin}%
\end{equation}
Using eq. (\ref{Fried}) and inserting the expression above into eq.
(\ref{FriedDot}) one gets%
\begin{equation}
\dot{H}+\frac{3\gamma}{2}H^{2}+\frac{k}{a^{2}}\left(  \frac{3\gamma}%
{2}-1\right)  =4\pi\gamma\rho_{\Lambda}\,. \label{tvlambda}%
\end{equation}
Once an expression for $\rho_{\Lambda}$ in terms of $H$ and/or $a$ is given,
this equation can be solved to obtain the time behavior of the scale factor.

\subsection{Viscous pressure \label{Bulk}}

The investigation of possible roles of dissipative processes in the
Universe has accompanied several developments of cosmology in the
past decades. The consideration of\ models with dissipation started
to raise considerable interest in the 1970's, both as a candidate to
explain the high entropy per baryon ratio inferred from the CMB
(see, e.g., ref. \cite{Weinberg71}), as well as a mechanism for
isotropization and homogenization of the Universe (see, e.g.,
\cite{CollinsStewart71}). Later, viscosity was invoked as a way to
avoid the primordial singularity (see, e.g., ref. \cite{Murphy73})
and to drive an inflationary expansion (see, e.g., refs.
\cite{WagaInflationary86,Barrow88,Gron90,Zimdahl96}). Its role in
the transition from a de Sitter epoch to the Friedmann epoch
(deflationary phase) was also investigated (see e.g.
\cite{WagaLimaPortugal,Zimdahl2000}). More recently, models with
bulk viscous pressure were investigated as possible sources for the
current phase of accelerated expansion of the Universe (see e.g.
refs. \cite{Zimdahl2001,Balakin2003}).

In the context of a homogeneous and isotropic Universe, dissipation
can only be present through a bulk viscous pressure (also known as
second viscosity), whose effect in the energy-momentum tensor (and
therefore, on the equations of
motion) is to add a new term to the isotropic dynamic pressure%
\begin{equation}
p=P-\Pi \, , \label{pPpi}%
\end{equation}
where $P$ is the equilibrium (thermostatic) pressure and $\Pi$ is a
correction term present in dissipative situations. This term can
appear either due to a real bulk viscosity --- to be derived from
first principles from kinetic theory (see, e.g., ref.
\cite{LL,Weinberg71}) --- or as an effective pressure associated to
the phenomenon of particle creation, which arises naturally in the
context of quantum processes (see e.g. \cite{Hu82}).

For first order deviations from equilibrium it may be shown that the
generic form of the extra term is given by
\cite{Eckart40,Weinberg71}
\begin{equation}
\Pi=\zeta\theta \, , \label{pilin}%
\end{equation}
where $\theta$ is the divergence of the four-velocitiy field
$u^\mu$, which in a homogeneous Universe, for comoving observers, is
simply given by $\theta=3H$. The viscosity coefficient $\zeta$ can
be a function of the dynamic variables ($\rho$, $p$, etc.), but not
of its (space-time) derivatives \cite{Eckart40,Weinberg71}. The
second law of thermodynamics requires that $\zeta>0$
\cite{Weinberg71}. On
phenomenological grounds, it is usually 
assumed that $\zeta$ is a function of the energy density $\rho$
only. For example, a generic power law form for the viscosity
coefficient%
\begin{equation}
\zeta=\alpha\rho^{n} \label{zBK}%
\end{equation}
was introduced and investigated in \cite{BelinskiiKhalatnikov77} and several
analytic solutions were discussed in \cite{Barrow88} and in subsequent papers.

As mentioned above, a bulk viscous pressure can also appear as a
consequence of quantum effects, such as particle production and
trace anomaly (for a calculation from first principles in the case
of a scalar field, see ref. \cite{Hu82}). An \textit{Ansatz} to
represent the effects of particle creation, which includes nonlinear
terms in the expansion factor, was given by Novello and Ara\'ujo
\cite{NovelloAraujo80}. In their work, the viscous term takes the
form,
\begin{equation}
\Pi=\sum_{k=1}^{N}\alpha_{k}\theta^{k}\,, \label{seriesTheta}
\end{equation}
where the coefficients can be functions of $\rho$.

Prigogini and collaborators \cite{PrigogineGRG89} applied the
thermodynamics of open systems to cosmology and derived the bulk
viscosity from particle creation. Assuming that the created
particles are in thermal equilibrium with the existing ones and that
the creation process occurs at constant specific entropy, they
found\footnote{This form of Prigogini et al.'s
result \cite{PrigogineGRG89} is taken from ref. \cite{CaLiWa92}.}%
\begin{equation}
\Pi=\frac{\rho+P}{n\theta}\Psi \, , \label{PiPrigo}%
\end{equation}
where $n$ is the particle number density and $\Psi$ is the particle
source (if $\Psi>0$) or sink (if $\Psi<0$), which is given by
$\Psi=N^{\mu}{}_{;\mu}$, where $N^{\mu}=n u^\mu$ is the particle
flux vector.
Calv\~{a}o et al. \cite{CaLiWa92} give a thermodynamical description
of particle creation processes in the Universe, still restricting to
adiabatic transformations, but relaxing the assumption of
constant specific entropy. They propose the \textit{Ansatz}%
\begin{equation}
\Pi=\frac{\beta\Psi}{\theta} \, , \label{PiCaLiWa}%
\end{equation}
which generalizes expression (\ref{PiPrigo}).

As in the previous sections, we shall consider only a fluid with the
equation of state\footnote{In models with $\gamma>2/3$, the bulk
viscous pressure can drive the accelerated expansion if $\Pi>\left(
\gamma-2/3\right)  \rho$. However, in this case the correction term
in (\ref{pPpi}) dominates over the thermodynamic pressure and is not
a small correction to the energy-momentum tensor. Therefore this
term is more naturally interpreted as arising from particle creation
processes.} $P=\left(  \gamma-1\right)  \rho$. Inserting expression
(\ref{pPpi}) in equation (\ref{FriedDot}) and using the Friedmann
equation (\ref{Fried}) with no cosmological term we obtain%
\begin{equation}
\dot{H}+\frac{3\gamma}{2}H^{2}+\frac{k}{a^{2}}\left(  \frac{3\gamma}%
{2}-1\right)  =4\pi\Pi \,. \label{viscosity}%
\end{equation}
A comparison with equation (\ref{tvlambda}) makes evident that the
viscous pressure can be equivalent to a cosmological term. If both
$\Pi$ and $\rho_{\Lambda}$ are expressed as functions of the scale
factor or the Hubble parameter alone, the explicit equivalence of
the two models is straightforward. For example, a (linear) bulk
viscosity of the form (\ref{pilin}) with constant viscosity
coefficient is equivalent to a cosmological term proportional to the
expansion factor: $\rho_{\Lambda }=(3\zeta/\gamma)H$. Therefore,
Sch\"{u}tzhold's model (eq. \ref{Hlin}) for $\Lambda$ is equivalent
to a linear viscosity with constant coefficient. However, in
general, if $\Pi$ and $\rho_{\Lambda}$ have different explicit
expressions in terms of $\rho$, $H$, or $a$, the equivalent models
can only be found after the solution of the equation of motion is
obtained.

\subsection{Nonlinear equations of state} 

Models with nonlinear equations of state have raised a substantial
interest recently, mainly due to the possibility of DE and DM
unification \cite{Tese,Bilicetal02}. For instance, if $p$ is a
nonlinear function of the energy density $\rho$, it could be
vanishing in high density regions, behaving as DM, and be close to
$-\rho$ in low density ones, acting as DE. This behavior would lead
naturally to a transition from matter domination to a de Sitter
state in the cosmic evolution and could have a minor impact in
processes taking place in the early Universe. Moreover, such
equation of state can lead to a positive (adiabatic) sound velocity
($c_{s}^{2}=dp/d\rho$), even for negative pressures.

An example of an equation of state exhibiting these properties is
given by an inverse power law
\cite{Tese,Kamen01,Bilicetal02}%
\begin{equation}
p=\frac{M^{4\left(  \alpha+1\right)  }}{\rho^{\alpha}} \,, \label{GCG}%
\end{equation}
where $M>0$ has dimension of mass and $\alpha$ is a real
dimensionless constant. The case $\alpha=1$ can be derived from the
Born-Infeld action for a scalar field \cite{BI} and arises from the
embedding of theories in higher dimensional space-times (in
particular from $(3+1)-$branes immersed in a $(4+1)-$bulk), being
connected with string theory \cite{Kamen01,dbranes,AppChap}. Because
of its similarity with an EOS proposed long ago in the context of
aerodynamics \cite{Chaplygin}, this fluid became know as Chaplygin
gas, and the generic power law (\ref{GCG}) as Generalized Chaplygin
Gas (GCG \cite{Kamen01,Bilicetal02}). This type of matter component
was dubbed as quartessence \cite{GCG1}, since in this framework,
only a fourth component, besides radiation, baryons, and neutrinos,
would be needed to describe the cosmic evolution.

The GCG model is consistent with a number of observational data
involving the background evolution of the Universe for
$-1/2\lesssim\alpha\lesssim1/2$ (see, e.g., refs.
\cite{GCG1,GCGtests}). On the other hand, equation (\ref{GCG}) fails
to reproduce LSS and CMB data if it is applied to the fluctuations
\cite{Sandviketal2004}. However, if the sound velocity vanishes (for
example, due to the introduction of a certain type of intrinsic
entropy perturbations), then the model is again in agreement with
the data \cite{nadiabGCG,silentGCG}.

Several other nonlinear EOS have been discussed in the literature
(see, for instance, refs.
\cite{Tese,ParkerRaval,ExpLog1,vdw,StepLike,MW,RevQuart}). Effective
equations of state appear generically in scalar field models with
noncanonical purely kinetic Lagrangians \cite{ScherrerK}. For
example, a generalization of the Born-Infeld action \cite{GCGk}
leads to equation (\ref{GCG}), while other generalizations yield new
classes of quartessence models (see, e.g., ref. \cite{EBI1}).

A natural generalization of equation (\ref{GCG}), in the context of
the linear EOS (\ref{plin}) considered in sections (\ref{Deacaying})
and (\ref{Bulk}) is
given by \cite{ModGCG,ModGCG0,ModGCG1,ModGCG2,SandroModGCG3}%
\begin{equation}
p=\left(  \gamma-1\right)  \rho-M^{4\left(  \alpha+1\right)  }\rho^{-\alpha
}\,, \label{linGCG}%
\end{equation}
which became know as the Modified Chaplygin Gas (MCG) EOS.

This type of EOS appears in several settings with different physical
motivations. For example, the consideration of Schwarzschild--Anti
de Sitter black holes in 5D \cite{KML2000} leads to the EOS
\begin{equation}
p=-\frac{2}{3}\rho-\frac{12}{\rho l^{2}}\,,%
\end{equation}
where $l$ is associated to the curvature scale of the asymptotic
anti-de Sitter space.

Another particular case of equation (\ref{linGCG}) that is useful in
the cosmological setting, is given by
\begin{equation}
p=\frac{\rho}{3}\left(1-\frac{\rho_{d}^{2}}{\rho^{2}}\right) \, ,
\label{relmass}
\end{equation}
which provides an excellent approximation for a relativistic gas of
massive particles \cite{ShaPeixSobr2004}.

For the purpose of illustrating the equivalence of the three
frameworks, it is convenient to consider more generic EOS with a
linear and a nonlinear part%
\begin{equation}
p=\left(  \gamma-1\right)  \rho-f\left(  \rho\right)  . \label{nlineos}%
\end{equation}
Inserting this expression into equation (\ref{FriedDot}) and using the
Friedmann equation (\ref{Fried}) with no cosmological term we obtain%
\begin{equation}
\dot{H}+\frac{3\gamma}{2}H^{2}+\frac{k}{a^{2}}\left(  \frac{3\gamma}%
{2}-1\right)  =4\pi f\left(  \rho\right)  \,. \label{frho00}%
\end{equation}

Clearly, the similarity of the equation above with equations
(\ref{tvlambda}) and (\ref{viscosity}) is manifest, showing that a
variable cosmic term, a bulk viscous pressure, and a nonlinear term
in the EOS can play an equivalent role in the determination of the
background cosmological dynamics. Although the aforementioned
equivalence between the three frameworks holds for generic equations
of state, we have restricted to fluids with a linear term to
simplify the discussion of some specific examples that shall be
developed in the next section.\\

It is worth noticing how the basic equations are modified in these
three classes of models. The introduction of a new matter-energy
component
--- the dynamical $\Lambda$ term --- alters both the energy
conservation and the Friedmann equations (eqs. \ref{econsL} and
\ref{Fried} respectively) with respect to a single fluid. On the
other hand, the inclusion of a viscous pressure modifies only the
energy conservation, through the inclusion of $\Pi$. Finally, a
nonlinear term in the EOS does not change any equation formally
(although it appears in both equations).

\section{Examples and analytic solutions} \label{results}

In this section we shall illustrate the connection among the three
roads to cosmic acceleration discussed above through some examples,
showing the explicit equivalence among specific models in each
framework. In particular, we shall consider two common types of
dynamical $\Lambda$ models: a power law cosmic term of the form
(\ref{Lpower}) and $\Lambda$-decaying models with a quadratic
dependence on the expansion rate, including models (\ref{H2}) and
(\ref{Hlin}) as particular cases. We explicitly find their analogues
in terms of a bulk viscous pressure and a nonlinear EOS in some
particular cases. In this process we derive novel analytical
solutions to these models.

\subsection{Cosmological term depending on the scale factor}\label{scale}

For models in which the $\Lambda$ term is given as a function of the
scale factor, the equivalence with a model with nonlinear EOS of the
form (\ref{nlineos}) can be easily tested if the energy conservation
equation allows an analytic solution, providing an explicit
expression for $\rho\left( a\right)$. In this case, $\rho
_{\Lambda}\left( a\right)$ is directly linked to the non-linear term
$f\left( \rho\left( a\right)\right)$.

As an example, let us consider the following EOS:
\begin{eqnarray}
p&=&\left(  \gamma-1\right)
\rho-\frac{B\rho}{1+\sqrt{1+C\rho}}\nonumber\\
&=&\left( \gamma-1\right) \rho+\frac{B}{C}\left(
1-\sqrt{1+C\rho}\right)
,\label{GCGmod1}%
\end{eqnarray}
where the parameter $B$ is dimensionless, and $C>0$ has dimensions
of inverse energy density. The solution of the energy conservation
equation (\ref{econsL}), for $\gamma\neq 0, %
$\footnote{The particular case $\gamma =0$ leads to the relation
$$\frac{\left(\sqrt{1+C\rho}-1\right)\exp\sqrt{1+C\rho}}
{\left(\sqrt{1+C\rho_0}-1\right)\exp\sqrt{1+C\rho_0}}
=\left(\frac{a}{a_0}\right)^{3B/2}\,.$$}
 is
\begin{equation}
\frac{\left( \frac{\sqrt{1+C\rho}+1-B/\gamma}{\sqrt{1+C\rho
_{0}}+1-B/\gamma}\right)
^{2\left(\frac{B-\gamma}{B-2\gamma}\right)}}{\left(
\frac{\sqrt{1+C\rho}-1}{\sqrt{1+C\rho_{0}}-1}\right)
^{\frac{2\gamma}{B-2\gamma}}}=\left( \frac{a}{a_{0}}\right)
^{-3\gamma}\,.\label{GCGmod1sol1}
\end{equation}

Choosing $B=\gamma$ this expression can be easily inverted, leading
to
\begin{equation}
\label{rhodea1} \rho=c_{1}a^{-3\gamma}+c_{2}a^{-3\gamma/2}\,,
\end{equation}
where $c_{1}=C^{-1}\left(  \sqrt{1+C\rho_{0}}-1\right)
^{2}a_{0}^{3\gamma}$ and $c_{2}=2\sqrt{c_{1}/C}$. Therefore, the
nonlinear term is given by
\begin{equation}
f\left(\rho\right)=\frac{\gamma}{C}\left(\sqrt{1+C\rho_0}-1\right)\left(\frac{a}{a_0}\right)^{-3\gamma/2}
=\gamma \frac{c_2}{2} a^{-3\gamma/2}\,. \label{nlrgcg}
\end{equation}
We thus see that this nonlinear term plays the same role as a
dynamical cosmic term of the form (\ref{Lpower}), with
$m=3\gamma/2$. Hence, in this case, the power law $\Lambda$ decaying
model\ with a $\gamma$-fluid (eq. \ref{plin}) is equivalent to a
single fluid with EOS (\ref{GCGmod1}), with $B=\gamma$. From
equation (\ref{nlrgcg}), it is clear that these models are also
analogous to a fluid with linear EOS and viscous pressure of the
form $\Pi=\Pi _{0}a^{-3\gamma/2}$ (see eq. \ref{viscosity}).

If on the one hand the evolution of the scale factor will be the
same in the three models, on the other hand analytical solutions may
be easier to find with one specific choice of the equivalent
descriptions. In particular, while it is not possible to find a
simple analytic solution for the energy conservation equation
(\ref{econsL}) with a power law decaying $\Lambda$ term and a linear
fluid, such solution is easily found for the nonlinear EOS
(\ref{GCGmod1}). The advantage of one description over the others in
deriving the evolution of the system will become even more apparent
in the discussion bellow.

Analytic solutions for $a(t)$ for the equivalent models discussed
above can be easily found for $\gamma=2n/3$ (where $n$ is an integer
number). For example, let us consider the ``radiation era''
($\gamma=4/3$), such that in the decaying $\Lambda$ model we have
$\rho_{\Lambda} \propto a^{-2}$ (or, in the case of viscous models,
$\Pi \propto a^{-2}$). In this case, equations (\ref{tvlambda}),
(\ref{viscosity}), and
(\ref{frho00}) allow the simple analytic solution%
\footnote{Here and throughout we shall impose the condition
$a(0)=0$. On the other hand, to simplify the expressions, the
``normalization'' of $a_0=a(t_0)$ will be left arbitrary.}
\begin{equation}
a=\left[  \left(  \frac{16\pi\sigma}{3}-k\right)  t^{2}+At\right]
^{1/2}\,,
\end{equation}
where $\sigma$ is the constant of proportionality present in each
model. The constant of integration $A$ is left free, since equations
(\ref{tvlambda}), (\ref{viscosity}), and (\ref{frho00}) are second
order differential equations (and we have already used $a(0)=0$).

The solution above is obtained in a simpler way for the nonlinear
fluid model, since in this case we have an explicit expression for
the energy density as a function of the scale factor (eq.
\ref{rhodea1}), which can be inserted in the Friedmann equation
(\ref{Fried}), without a cosmological term, yielding
\begin{equation}
\label{H2c1c2}
H^{2}=\frac{8\pi}{3}\left(c_{1}a^{-3\gamma}+c_{2}a^{-3\gamma/2}\right)
-\frac{k}{a^{2}}\,.
\end{equation}
For $\gamma=4/3$ this equation becomes
\begin{equation}
H^{2}=\frac{8\pi c_{1}}{3}a^{-4} -\left(k-\frac{8\pi
c_2}{3}\right)a^{-2}\,,
\end{equation}
which is analogous to the one obtained for a Universe with only
radiation and curvature, with the quantity $k-8\pi c_2/3$ in the
place of the curvature parameter $k$. Now we have a first order
differential equation which can be easily solved to get
\begin{equation}
a=\left[  \left(  \frac{8\pi c_2}{3}-k\right)
t^{2}+2t\sqrt{\frac{8\pi c_1}{3}}\right] ^{1/2}\,.
\end{equation}
Now the constant $A$ has a clear interpretation in terms of the EOS
parameters and the ``normalization'' $\rho(a_0)=\rho_0$.

A simple analytic solution can also be found from eq. (\ref{H2c1c2})
in the case $\gamma=2/3$:
\begin{equation}
a=\frac{2\pi c_2}{3}t^2+t\sqrt{\frac{8\pi c_1}{3}-k}\,.
\end{equation}
Of course, this solution is also valid for a cosmological term
decaying as $a^{-1}$ in a Universe dominated by a fluid with EOS
$p=-\rho/3$ (e.g., a gas of nonrelativistic cosmic strings) for any
curvature $k$.

These simple examples illustrate how the use of equivalent models is
handy for obtaining analytic solutions. In particular, for decaying
$\Lambda$ models where $\rho_{\Lambda}$ is a function of the scale
factor $a$, it may be more advantageous to use an equivalent model
in terms of a nonlinear fluid, since it may allow to obtain analytic
solutions for $\rho(a)$.

\subsection{Cosmological term depending on the expansion rate} \label{HLambda}

We shall now consider the case where the cosmic term is expressed as
a function of $H$. More specifically, we introduce the {\it Ansatz}
\begin{equation}
\Lambda=\Lambda_{0}+\sigma H+\nu H^{2}\,, \label{quadL}
\end{equation}
which can be regarded as a series expansion of a generic
$\Lambda(H)$.

Clearly, from equations (\ref{tvlambda}) and (\ref{viscosity}), it
is easy to see that the dynamics of a Universe with a cosmological
term of the form above and filled by a perfect fluid is equivalent
to the one of a Universe with no cosmic term, but with and imperfect
fluid with viscous pressure of the form (\ref{seriesTheta}) with
constant coefficients (and including a constant term).

On the other hand, to make the comparison with nonlinear fluid
models --- for generic curvature --- the combination of the
Friedmann and energy conservation equations has to be used. Some
examples that admit analytic solutions for $a(t)$, are discussed
below.

\subsubsection{Trace anomaly cosmological term}\label{Trace}

Let us first consider Sch\"{u}tzhold's model \cite{Schutzhold02} of
decaying $\Lambda$ (eq. \ref{Hlin}), which has the form
(\ref{quadL}) with $\Lambda_0=0$ and $\nu=0$. As mentioned in
section \ref{Bulk}, in the case of a linear fluid (eq.~\ref{plin}),
this model is equivalent to a viscous fluid with constant viscosity
coefficient.

In a flat Universe, this model leads to the analytic solution
\cite{TreciokasEllis,Gron90,BorgesSaulo05}
\begin{equation}
a=\left[\exp\left(\sigma\gamma t/2
\right)-1\right]^{\frac{2}{3\gamma}}\,.\label{Hk0}
\end{equation}

As pointed out by S. Carneiro \cite{SauloEq}, this same solution is
found in the case of the MCG (eq. \ref{linGCG}) with $\alpha=-1/2$.
In fact the energy conservation equation allows a
simple analytic solution for this type of fluid%
\footnote{For $\gamma=0$ the solution is
$$\rho=\rho_0\left[1+3\left(1+\alpha\right)
\left(\frac{M^4}{\rho_0}\right)^{\left(1+\alpha\right)}\ln\frac{a}{a_0}\right]^{\frac{1}{1+\alpha}}\,.$$}:
\begin{equation}
\rho=\rho_{0}\left[  \left(  1-\frac{A}{\gamma}\right)
a^{-3\gamma\left( 1+\alpha\right)  }+\frac{A}{\gamma}\right]
^{\frac{1}{1+\alpha}}\,,
\end{equation}
where $\rho_{0}$ is a constant of integration and $A=\left(
M^{4}/\rho _{0}\right)  ^{1+\alpha}$. %
Thus, in the flat case and for $\alpha=-1/2$, the nonlinear term is
given by
\begin{equation}
f\left(\rho\right)=M^2\rho^{1/2}=HM^2\sqrt{\frac{8\pi}{3}}\,,
\end{equation}
showing the equivalence of the two models.

For some values of $\gamma$, this nonlinear fluid model with
$\alpha=-1/2$ allows analytic solutions for any value of the
curvature $k$. For example, if $\gamma=4/3$ (which can be regarded
as a ``generalized radiation fluid'') one has,
\begin{equation}
a = \left[
e^{\sqrt{6\pi}M^2t}-1-\frac{k}{\frac{4\sqrt{\rho_{0}}}{3M^2}-1}\left(
\frac{\sinh\sqrt{\frac{3\pi}{2}}M^2t}
{\sqrt{\frac{3\pi}{2}}M^2}\right) ^{2}\right]
^{\frac{1}{2}},\label{GRF}
\end{equation}
while for $\gamma=2/3$ (``generalized cosmic string gas'') one has
\begin{eqnarray}
\label{a23} a &=& 2\pi \rho^{1/2}_0
M^2\left(1-\frac{3M^2}{2\rho_0^{1/2}}\right)\left(\frac{\sinh\sqrt{\frac{3\pi}{2}}M^2t}
{\sqrt{\frac{3\pi}{2}}M^2}\right)^2\nonumber\\
&&+\left[\frac{8\pi
\rho_0}{3}\left(1-\frac{3M^2}{2\rho_0^{1/2}}\right)^2-k\right]^{1/2}\frac{\sinh\sqrt{6\pi}M^2t}{\sqrt{6\pi}M^2}
\,.\nonumber\\ \label{GCSG}
\end{eqnarray}

These new solutions generalize equation (\ref{Hk0}), for these two
choices of $\gamma$, for arbitrary curvature. 
Although these solutions are not directly related to a constant bulk
viscosity or cosmic term proportional to $H$, they exemplify how the
use of this nonlinear fluid model, by analytically solving the
energy conservation equation, simplifies the obtention of analytic
solutions for $a(t)$.

Analytic solutions for $\alpha=1$ and $\gamma = 4/3$, in the flat
case (i.e., for EOS \ref{relmass}), were obtained in reference
\cite{ShaPeixSobr2004}. It is worth mentioning that this model is
equivalent (again in the flat case) to a linear fluid with particle
creation, leading to a bulk viscous term of the form
(\ref{PiCaLiWa}) with constant source and $\beta \propto
\rho^{-1/2}$. On the other hand, if $\beta = const.$, this particle
creation model is equivalent to a MCG with $\alpha=1/2$.

\subsubsection{Renormalization group cosmological term} \label{RGL}

Another interesting case is found for $\sigma=0$ in equation
(\ref{quadL}), which gives the form (\ref{H2}) representing the
effect of quantum corrections to the vacuum energy in the
renormalization group approach
\cite{ShapiroetalPLB03,BonetetalJCAP04,ShaSoSteJCAP05}. In the flat
case, the effect of this cosmic term is clearly analogous to having
a term $f(\rho)=(\Lambda_0/8\pi)+(\gamma \nu / 3)\rho$ in equation
(\ref{frho00}). This analogy is still valid for generic curvature,
as can be seen by inserting expression (\ref{H2}) into equation
(\ref{tvlambda}):
\begin{equation}
\label{Hdotbeta}
\dot{H}+\frac{\gamma}{2}\left(3-\nu\right)H^2+\frac{k}{a^2}\left(\frac{3\gamma}{2}-1\right)
=\frac{\gamma\Lambda_0}{2}\,.
\end{equation}
This equation has the same solution as for a model with constant
$\Lambda$ and linear EOS of the form (\ref{plin}), but with
$\gamma^\prime=\gamma (1-\nu/3)$, $k^\prime=k
(3\gamma/2-1)/(\gamma(3-\nu)/2-1)$, and
$\Lambda_0^\prime=\Lambda_0/(1-\nu/3)$.

The case $\gamma=1/\left(3-\nu\right)\,$ (i.e.,
$\gamma^\prime=1/3$), allows an analytic solution given by
\begin{eqnarray}
a\left(t\right)&=&\sinh^2\left[\frac{t}{2}\sqrt{\frac{\Lambda_0}{3-\nu}}\right]\nonumber\\
&&+\left[\frac{k}{\Lambda_0}\left(2\nu
-3\right)\right]^{\frac{1}{2}}\left[e^{t\sqrt{\Lambda_0/\left(3-\nu\right)}}-1\right]\,.
\label{sigma0}
\end{eqnarray}

An explicit solution for $\gamma=4/(3-\nu)$ can also be obtained as
a particular case of equation (\ref{quadSolK}), presented in the
next section.

\subsubsection{Quadratic expansion rate cosmological term}

As a final example, let us consider the full quadratic form (eq.
\ref{quadL}). Inserting this expression in equation (\ref{tvlambda})
and introducing the new variable $u$, such that $a=u^n$, with
\begin{equation}
n=
\frac{2}{\gamma\left(3-\nu\right)}\,,
\end{equation}
one has
\begin{equation}
\ddot{u}-\frac{\gamma\sigma}{2}\dot{u}-\frac{\gamma\Lambda_0}{2n}u=-\frac{k}{n}
\left(\frac{3\gamma}{2}-1\right)u^{1-2n}\,.
\end{equation}

For $k=0$ (or $\gamma=2/3$), the solution is
\begin{eqnarray}
\label{uH} u=u_H&:=&e^{\gamma\sigma t/4}
\left\{c_1\exp\left[t\left(\frac{\gamma^2\sigma^2}{16}+\frac{\gamma\Lambda_0}{2n}
\right)^{\frac12}\right]\right.\nonumber\\
&&\left.+c_2\exp\left[-t\left(\frac{\gamma^2\sigma^2}{16}+\frac{\gamma\Lambda_0}{2n}\right)
^{\frac12}\right]\right\}\,, \label{quadSolK0}
\end{eqnarray}
where $c_1$ and $c_2$ are arbitrary constants.

Imposing $a(0)=0$ as in the previous cases, we have
\begin{eqnarray}
a(t)=\left\{e^{\gamma\sigma t/4}
\sinh\left[t\left(\frac{\gamma^2\sigma^2}{16}+\frac{\gamma\Lambda_0}{2n}
\right)^{\frac12}\right]\right\}^n\, . \label{sinh}
\end{eqnarray}
Notice that for $\nu=\Lambda_0=0$ we recover the solution
(\ref{Hk0}), as it should be. Also, for $\sigma=0$ and
$\gamma=1/(3-\nu)$, we recover expression (\ref{sigma0}) for $k=0$.

Solutions for $k\neq 0$ can be found in some particular cases. For
example, for ~$n=1/2$, such that $\gamma=4/(3-\nu)$, the solution is
\begin{equation}
u=u_H+k\left( \frac{3\gamma}{2}-1\right)\frac{2}{\gamma \Lambda_0}
\,.
\end{equation}
Again, imposing $a(0)=0$, we have
\begin{eqnarray}
&&a\left(t\right)\!=\!\left\{e^{\gamma\sigma
t/4}\sinh\left[t\left(\frac{\gamma^2\sigma^2}{16}+
\gamma\Lambda_0\right)^{\frac12}\right] \right.\nonumber\\&&\left.
\!\!+\!\frac{2k}{\gamma\Lambda_0}\!\!\left(\!\frac{3\gamma}{2}\!-\!1\!\right)\!\!\!\left(\!1\!-\!e^{\gamma\sigma
t/4}\!\cosh\!\left[\!t\!\left(\!\frac{\gamma^2\sigma^2}{16}\!+\!\gamma\Lambda_0\!\!\right)^{\!\frac12}\!\right]
\right)\!\!\right\}^{\!\frac12}. \label{quadSolK}
\end{eqnarray}

A nonlinear fluid model that leads to the same solutions as the
quadratic $\Lambda(H)$ model, in some cases, 
is given by the
following EOS
\begin{equation}
p=\left(\gamma-1\right)\rho-M^2\rho^{1/2}-A\,, \label{GCGc}
\end{equation}
with $A$ being a constant. This model allows an analytic solution
for the energy conservation equation, which, for $\gamma\neq 0$, is
given by\footnote{For $\gamma=0$ the relation is
$$\rho^{1/2}-\rho_0^{1/2}-\frac{A}{M^2}\ln\left(\frac{M^2\rho^{1/2}+A}{M^2\rho_0^{1/2}+A}\right)=\frac{3M^2}{2}\ln\frac{a}{a_0}\,.$$}
\begin{equation}
\left(\frac{a}{a_0}\right)^{-3\gamma}=%
\frac{%
\left(\frac{\rho^{1/2}-\frac{M^2}{2\gamma}-\sqrt{\frac{M^4}{4\gamma^2}+\frac{A}{\gamma}}}
{\rho^{1/2}_0-\frac{M^2}{2\gamma}-\sqrt{\frac{M^4}{4\gamma^2}+\frac{A}{\gamma}}}\right)^{\frac{M^2}{\sqrt{M^4+4A\gamma}}+1}}%
{
\left(\frac{\rho^{1/2}-\frac{M^2}{2\gamma}+\sqrt{\frac{M^4}{4\gamma^2}+\frac{A}{\gamma}}}
{\rho^{1/2}_0-\frac{M^2}{2\gamma}+\sqrt{\frac{M^4}{4\gamma^2}+\frac{A}{\gamma}}}\right)^{\frac{M^2}{\sqrt{M^4+4A\gamma}}-1}}
 \,.
 \label{arho}
\end{equation}

Clearly, for $k=0$ the model above (eq. \ref{GCGc}) is equivalent to
the quadratic $\Lambda$ model (eq. \ref{quadL}), and has a solution
of the form (\ref{sinh}). However, this solution would hardly be
found from inverting relation (\ref{arho}) and inserting the result
in the Friedmann equation. This is an example where the use of a
nonlinear EOS makes it harder to obtain analytic solutions, contrary
to the examples of sections \ref{scale} and \ref{Trace}.

\section{Summary} \label{summary}

In this work it was shown that three distinct classes of
cosmological models (decaying $\Lambda$, bulk viscous pressure and
fluids with exotic nonlinear equations of state) are equivalent, in
the sense that in each of the classes there are models that
reproduce exactly the same expansion history of the Universe. This
equivalence is apparent from the formal analogy among equations
(\ref{tvlambda}), (\ref{viscosity}), and (\ref{frho00}).%
\footnote{It is worth mentioning that the equivalence among these
frameworks has been discussed in previous works for some particular
cases. For example, the equivalence between variable-$\Lambda$
cosmological models and models with bulk viscosity was pointed out
in refs. \cite{Berman} (for one specific model in the flat case) and
\cite{Beesham} (for two models). Also, the similarity of the trace
anomaly cosmic term with the MCG with $\alpha=-1/2$, in the flat
case, was shown in ref. \cite{SauloEq}.}

As explicit examples where the equivalence is manifest, we have
considered two popular classes of ``decaying vacuum'' cosmologies: a
power law $\Lambda$ term (sec. \ref{scale}) and a cosmic term
depending on the expansion rate (sec. \ref{HLambda}).

We have analyzed the case of a generic quadratic cosmological term
of the form $\Lambda=\Lambda_{0}+\sigma H+\nu H^{2}$, which is
equivalent to a viscous term of the form (\ref{seriesTheta}) with
$N=2$ and constant coefficients, plus a constant term. In the flat
case (and for a linear EOS $p=(\gamma-1)\rho$) this model is also
equivalent to the MCG Gas with $\alpha=-1/2$ plus a constant term
(eq. \ref{GCGc}).

We obtained a complete analytical solution for these models in the
flat case (eq. \ref{quadSolK0}). For arbitrary curvature, we have
shown a solution for $\gamma = 4/(3-\nu)$ (eq. \ref{quadSolK}). As
far as we know, it is the first time that such type of model is
discussed in the literature. 

For $\Lambda_{0}=0$, and in the flat case, this model is equivalent
to a linear viscosity of the form (\ref{pilin}) with coefficient
$\zeta=\zeta_0+\zeta_1 \rho^{1/2}$ and, for a linear EOS, to the MCG
(eq. \ref{linGCG}) with $\alpha=-1/2$.

For $\Lambda_{0}=0=\nu$ we recover the ``trace anomaly cosmological
term'', which is equivalent to a linear viscous term with constant
viscosity coefficient. Again, in the flat case and for a linear EOS,
this model is equivalent to the MCG with $\alpha=-1/2$.

We have found two analytic solutions for the MCG with $\alpha=-1/2$
and arbitrary curvature: one for the ``generalized radiation fluid''
(eq.~\ref{GRF}) and another for the ``generalized cosmic string
gas'' (eq.~\ref{GCSG}). This generalizes
the solution for the flat case 
discussed in \cite{BorgesSaulo05}.

For $\sigma=0$ we recover the ``renormalization group cosmic term''.
We have show that this $\Lambda$-decaying model is equivalent to a
model with constant $\Lambda$ and a linear fluid, with redefined
values for the cosmological constant, the curvature, and the
equation of state parameter (see sec. \ref{RGL}). We have obtained
an analytic solution for the case $\gamma=1/(3-\nu)$ and arbitrary
curvature.

Regarding the power law $\Lambda$ term, we found that models with a
linear equation of state $p=(\gamma-1)\rho$ and a power law cosmic
term of the form $\Lambda \propto a^{-3\gamma/2}$ (or equivalently a
viscous pressure of the form $\Pi = \Pi_0 a^{-3\gamma/2}$) are
equivalent to a model with EOS given by $p=(\gamma-1)\rho+\gamma
C^{-1}\left(1-\sqrt{1+C\rho}\right)$ and no $\Lambda$. We have
displayed the analytical solutions for $a(t)$ for $\gamma=4/3$ and
$\gamma=2/3$, for generic curvature.

The use of equivalent models can be handy in finding analytic
solutions. In fact, in some cases one particular model excels the
other ones in the task of deriving explicit solutions. Here, we have
illustrated this property with particular examples where the 
solutions to a particular model are easily found using the
equivalent model belonging to another class.

\section{Discussion and concluding remarks \label{discussion}}

As mentioned above, a practical application of the mathematical
equivalence pointed out in this work is in the search for analytical
solutions, which can be useful for pedagogical purposes and to test
numerical codes. But, besides this operational use, does the
equivalence among the three classes of models have a deeper physical
connection? Can we observationally distinguish these models?

Certainly they cannot be distinguished with observables that are
mostly dependent on the background cosmology, such as the
redshift-distance diagram of type Ia supernovae. On the other hand,
although their dynamic role is equivalent, these models may have
different thermodynamical behavior. For example, it is know that
bulk viscosity and matter creation processes are thermodynamically
distinct (see, e.g. \cite{LimaGermano})%
\footnote{The thermodynamics of a vacuum decaying model is discussed
by Alcaniz \& Lima \cite{AlcanizLimaPRD72}, whereas the
thermodynamic behavior of the GCG was investigated by Santos,
Bedran, \& Soares \cite{Bedran}.}. Thus, one may try to use
thermodynamic arguments to distinguish among the three frameworks.

The existence of equivalent models in the three classes discussed in
this work is not very surprising given the limited degrees of
freedom of a homogeneous Universe. In fact a similar equivalence was
pointed out in refs. \cite{ChibaNakamura} connecting scalar field
models with nonlinear fluids in the homogeneous case.

A way to distinguish these models is to study their behavior when
more degrees of freedom are present. Conversely, if the equivalence
among the three frameworks remains in more generic configurations,
this could point to a deeper connection among them. Thus, the
natural next step in this investigation is to study the behavior of
linear perturbations in the three classes of models.

Naturally, if $\Lambda$ decaying models are considered homogeneous
by construction, the perturbations of the inhomogeneous component
will be clearly different from viscous models or models with
nonlinear equations of state, since the viscous or nonlinear terms
are intrinsically inhomogeneous. On the other hand, the vacuum
contribution can be regarded being as space-time dependent. For
example, in models where the cosmic term depends on the expansion
rate, $H$ can be replaced by a third of the local value of the
velocity divergence (see, e.g., \cite{FSS4v}).

Models with bulk viscous pressure are readily applicable to
inhomogeneous situations by using the local values of the density
and velocity divergence in the expression of the viscous term. Of
course, models which are identical in the flat case for homogeneous
configurations will have a different behavior in the perturbations.

It is also straightforward to consider adiabatic perturbations in
models with nonlinear equations of state. However, for quartessence
models such as the GCG, this type of perturbations generate large
scale structures that are inconsistent with observational data (see,
e.g, \cite{Sandviketal2004,ExpLog1}). On the other hand, a specific
type of entropy perturbation brings these models back in agreement
with the data \cite{nadiabGCG,ExpLog1,silentGCG}. Several other ways
to generalize to the inhomogeneous case equations of state that
correspond to the GCG for the background have also been considered
(see, e.g., refs. \cite{JulioPert,BertolamiSep,KR4ess}).

The study of perturbations in the three frameworks is rather
involving and presents several subtleties. Therefore, this analysis
is left for subsequent works (see, e.g., \cite{SandroPert}).

It is also worth investigating the microphysical motivation of the
three frameworks. For example, it is known that models with
nonlinear equations of state may arise from scalar fields with
noncanonical kinetic terms \cite{ScherrerK,EBI1,GCGk}.

Another avenue of research is to investigate wether the equivalence
holds with respect to other frameworks for cosmic acceleration, such
as self-interacting gases \cite{Zimdahl2001,Balakin2003} and other
dark energy models.

Finally, several cosmological observables (such as large-scale
structure, cosmic microwave background fluctuations, and
distance-redshift relations) have to be obtained in order to check
the viability of these models with respect to the diverse set of
observational data available.

Once these steps are undertaken, the study of equivalent models may
prove useful for finding physical insights and motivations to
phenomenological and first principle models that provide a good
description of the available astro-cosmological data.

\section*{Acknowledgements}

We thank Saulo Carneiro for discussions that inspired this work and
the organizers of the \textit{Workshop Nova F\'{\i}sica no
Espa\c{c}o}, \textit{III Workshop Brasileiro em Gravita\c{c}\~{a}o e
Cosmologia}, and of the \textit{100 years of relativity}\
conference, where this work was discussed. SSC thanks FAPEMAT for
financial support and ICRA-CBPF for their hospitality. MM thanks the
organizers of the \textit{2$^a$} \textit{Escola Mato-grossense de
F\'{\i}sica} and the hospitality of UFMT.

\end{document}